\documentclass[trackchanges, twocolumn]{aastex7}

\usepackage{hyperref}

\usepackage{comment}

\newcommand{\chandra}{{Chandra}}

\newcommand{\xmm}{{XMM-Newton}}

\newcommand{\nustar}{{NuSTAR}}

\newcommand{\fluxcgs}{\ensuremath{\mathrm{erg}\,\mathrm{s}^{-1}\,\mathrm{cm}^{-2}}}
\newcommand{\lumcgs}{\ensuremath{\mathrm{erg}\,\mathrm{s}^{-1}}}

\begin{document}

\title{Synchrotron radiation from NGC 470 HLX1 - a hidden hyperluminous accreting neutron star?}

\author[orcid=0000-0002-3033-5843]{Tanuman Ghosh}
\affiliation{Inter-University Centre for Astronomy and Astrophysics, Post Bag 4, Ganeshkhind, Pune 411007, India}
\email[show]{tanuman.ghosh@iucaa.in}

\author{Shiv Sethi}
\affiliation{Astronomy and Astrophysics, Raman Research Institute, C. V. Raman Avenue, Sadashivanagar, Bangalore 560080, India}
\email{sethi@rri.res.in}

\author{Gulab Chand Dewangan}
\affiliation{Inter-University Centre for Astronomy and Astrophysics, Post Bag 4, Ganeshkhind, Pune 411007, India}
\email{gulabd@iucaa.in}

\author{Matteo Bachetti}
\affiliation{Istituto Nazionale di Astrofisica-Osservatorio Astronomico di Cagliari, via della Scienza 5, I-09047 Selargius (CA), Italy}
\email{matteo.bachetti@inaf.it}

\author{Vikram Rana}
\affiliation{Astronomy and Astrophysics, Raman Research Institute, C. V. Raman Avenue, Sadashivanagar, Bangalore 560080, India}
\email{vrana@rri.res.in}

\author{Ranjeev Misra}
\affiliation{Inter-University Centre for Astronomy and Astrophysics, Post Bag 4, Ganeshkhind, Pune 411007, India}
\email{rmisra@iucaa.in}

\begin{abstract}
We present the first broadband spectral analysis of NGC 470 HLX1, a hyperluminous X-ray source that exhibits significant flux variability over different epochs. We investigate the feasibility of synchrotron radiation with varying latitude from a magnetized neutron star to explain the source's spectra. Although the statistical quality of the data does not allow us to rule out simple phenomenological models, it is adequate to constrain the parameter space for this more physically motivated model. We also shed light on the particle acceleration mechanisms and maximum Lorentz factor of electrons within the neutron star magnetospheric plasma under super-Eddington accretion conditions. In our broadband spectral modeling, the detection of a blackbody-like component suggests the presence of a disk near the corotation radius or an outflow ejected from the disk. The viability of synchrotron emission in an HLX system offers new insights about the nature of these sources, motivating further sample studies to assess whether most of these sources are powered by accreting neutron stars. 

\end{abstract}



\section{Introduction}
Hyperluminous X-ray sources (HLXs) are a subclass of ultraluminous X-ray sources (ULXs) with intrinsic X-ray luminosity ($L_x$) exceeding $10^{41}$ \lumcgs (see e.g., \citealt{Kaaret2017, King2023}). Although the X-ray spectral and timing properties of HLXs and ULXs are not very different, due to the extreme luminosity of HLXs, they are still best understood to be potential intermediate-mass black hole (IMBH) candidates (e.g., \citealt{Gao2003, Farrell2011, Barrows2019}). In contrast, from modern observational evidence, most of the ULXs have been strongly perceived to be stellar-mass compact objects with super-Eddington accretion. The discovery of ULX pulsar confirmed the notion of a neutron star as the powerhouse of these sources (e.g., \citealt{Bachetti2014}). Especially, a source NGC 5907 ULX1, which had shown maximum luminosity $> 10^{41}$ \lumcgs, an ideal IMBH candidate by the perception of HLXs, turns out to be a pulsar \citep{Israel2017}. In fact, there was a significant flux drop in this source over time, which was assumed to be correlated with the propeller phase transition \citep{Furst2023}, as it is understood to be highly probable for other extreme ULX pulsars as well.

In terms of broadband X-ray spectral modeling of ULXs, typically a three-component model is required, which involves cool and hot thermal components coming from wind and disk, respectively, and a phenomenological cutoff powerlaw model, which takes care of the high-energy spectral curvature ($\sim 10$ keV) typically found in most of the ULXs (e.g., \citealt{Walton2018}). Several interpretations of the high energy cutoff involve external down-scattering from the corona in a stellar-mass black hole scenario or emission from the column for a highly magnetized neutron star (e.g., \citealt{Walton2020}). Recently, we developed another model that shows synchrotron emission with a high latitude angle can explain this $\sim 10$ keV cutoff \citep{Ghosh2023}.

In \citealt{Ghosh2023}, mainly, the feasibility of the physical process and possible parameters were discussed in the expected scenario of a highly magnetized neutron star for two case studies of known pulsar ULXs (PULXs), namely, NGC 5907 ULX1, and NGC 7793 P13.  It is also prudent to further shed light on the location of radiation and particle acceleration processes responsible for triggering highly relativistic particles to emit synchrotron photons in this context. 

In literature, several electron acceleration mechanism have been explored relevant to accreting neutron star systems, such as collisional shock acceleration due to wind from the companion onto the Alfven radius of the neutron star \citep{Bednarek2009}, large-amplitude Deutsch waves \citep{Gunn1969, Eichler1984}, acceleration via electromotive forces, and magnetic reconnection due to strong conduction currents in the disk-magnetosphere-star circuit \citep{Miller1994, Hamilton1994}, and magnetospheric gap induced acceleration \citep{Cheng1989, Cheng1991, Zhang2014}. Based on those acceleration mechanisms, the feasibility of a non-thermal synchrotron is evident in accreting neutron star systems.

In this paper, we discuss the first broadband X-ray analysis of an extremely luminous ULX (peaked luminosity reached $\sim 1.5 \times 10^{41}$ \lumcgs, hence an HLX; \citealt{Walton2011, Sutton2012}), NGC 470 HLX1 (distance $D \sim 34$ Mpc), by exploring the high-energy spectrum as the synchrotron emission component along with soft emission counterparts. The source was previously observed by \chandra\ and \xmm\ and analyzed by \citealt{Sutton2012}. One of the primary implications regarding the source, similar to other HLXs, is that the source might host an IMBH candidate since those archival X-ray spectra did not show typical ULX-like spectral curvature or soft excess, possibly due to low count statistics, and the absence of hard X-ray data. In this work, we explore the possibilities beyond those previous claims, as discussed below.

First, we briefly discuss and recapitulate the synchrotron model and some of its theoretical implications in Section \ref{Sec: Synchrotron Model}. In Section \ref{Sec: Data}, we describe the data analysis method. Thereafter, we present the first-ever modeling of the broadband spectra with the newly developed synchrotron model in NGC 470 HLX1 in Section \ref{Sec: Results}. In section \ref{sec:discussions}, we discuss the feasibility of the results, the parameter space, and possible acceleration mechanisms in these systems, and try to understand the physical scenario of this particular HLX.

\section{Recap of the synchrotron model} \label{Sec: Synchrotron Model}
This paper primarily explains the high-energy spectrum with the high-latitude synchrotron emission model proposed in \citealt{Ghosh2023}. Here we briefly recapitulate the main aspects of the model. The radiated power in the $n$-th harmonic for a single electron can be expressed as:

\begin{equation}
  \small dI_n / d\Omega = \frac{e^2\omega^2}{2\pi c} \left[\tan^2\theta J^2_n(n\beta\cos\theta)  +\beta^2 J'^2_n(n\beta\cos\theta) \right]
  \label{eq:synrad_ang}
\end{equation}

where $\beta = v/c$, $v$ is velocity of the electron, $c$ is speed of light in vacuum. The integer $n$ identifies the harmonic for which $\omega = n\,\omega_B$, with $\omega_B = eB/(\gamma m_e c)$. The relativistic Lorentz factor is $\gamma = 1/\sqrt{1-\beta^2}$, and $\omega$ corresponds to the angular frequency of the photons. $B$ denotes the magnetic field strength, and $\theta$ is the angle between the direction of the emitted radiation and the orbital plane of the particle. The quantities $J_n(x)$ and $J_n'(x)$ represent the Bessel function of order $n$ and its derivative, respectively. (also see \citealt{Ghosh2023}). $d\Omega = dA/D^2$, where $D$ is the distance to the source and $dA$ is the area collected at the detector when observed. To get the power per unit frequency band, in a continuous distribution of frequencies, we know that $dI/d\omega = I_n/\omega_B$ \citep{Landau, Schwinger}. Thus, the final flux spectrum is obtained after integrating over the electrons' energy distribution:

\begin{equation}
 \small  F_\nu = N'
  \int_{\gamma_{\rm min}}^{\gamma_{\rm max}} S_\nu \gamma^{-p} \exp\left(-\gamma/\gamma_{\rm max}\right) d\gamma , 
  \label{eq:total_spec}
\end{equation}

\begin{equation}
 \small N' = \frac{N}{\int_{\gamma_{\rm min}}^{\gamma_{\rm max}} \gamma^{-p} \exp\left(-\gamma/\gamma_{\rm max}\right) d\gamma}
\end{equation}
and,
\begin{equation}
 \scriptsize S_\nu = \frac{2\pi e^2\nu^2}{c\nu_B} \left[\tan^2\theta J^2_\frac{\nu}{\nu_B}\left(\frac{\nu}{\nu_B}\beta \cos\theta\right)  + \beta^2 J'^2_\frac{\nu}{\nu_B}\left(\frac{\nu}{\nu_B} \beta \cos\theta\right) \right]
\end{equation}

$N = \rho V /D^2$, $\rho$ is the relativistic particle number density responsible for the emission, $V$ is the emitting region volume, and $D$ is the distance to the source.  We assume $\gamma_{max} = 1000 \gamma_{min}$, $p=2.2$, following the prescription of \citealt{Ghosh2023}.

Here we note that in equation 6 of \citealt{Ghosh2023}, we have used $\beta=1$ in the final model, which applies to a wide range of parameters. However, here we use the final spectrum derived exactly from Eq. \ref{eq:synrad_ang}, without assuming $\beta=1$, for completeness, which includes a broader range of parameter space. This choice of $\beta$ does not alter any implications to the spectral fittings/results obtained in \citealt{Ghosh2023}. Also, instead of $\gamma_{min}/B$ as a single parameter, we now separate them as two parameters (also due to the usage of non-unity of $\beta$). Although we reaffirm here that the degeneracy of $\gamma$ and $B$ is intrinsic to the equation for high latitude emission, hence, eventually, we can only constrain two parameters (namely $N$ and $\gamma_{min}$) for fixed $B$ and $\theta$, similar to \citealt{Ghosh2023}.

\begin{figure}
    \centering
    \includegraphics[width=\linewidth]{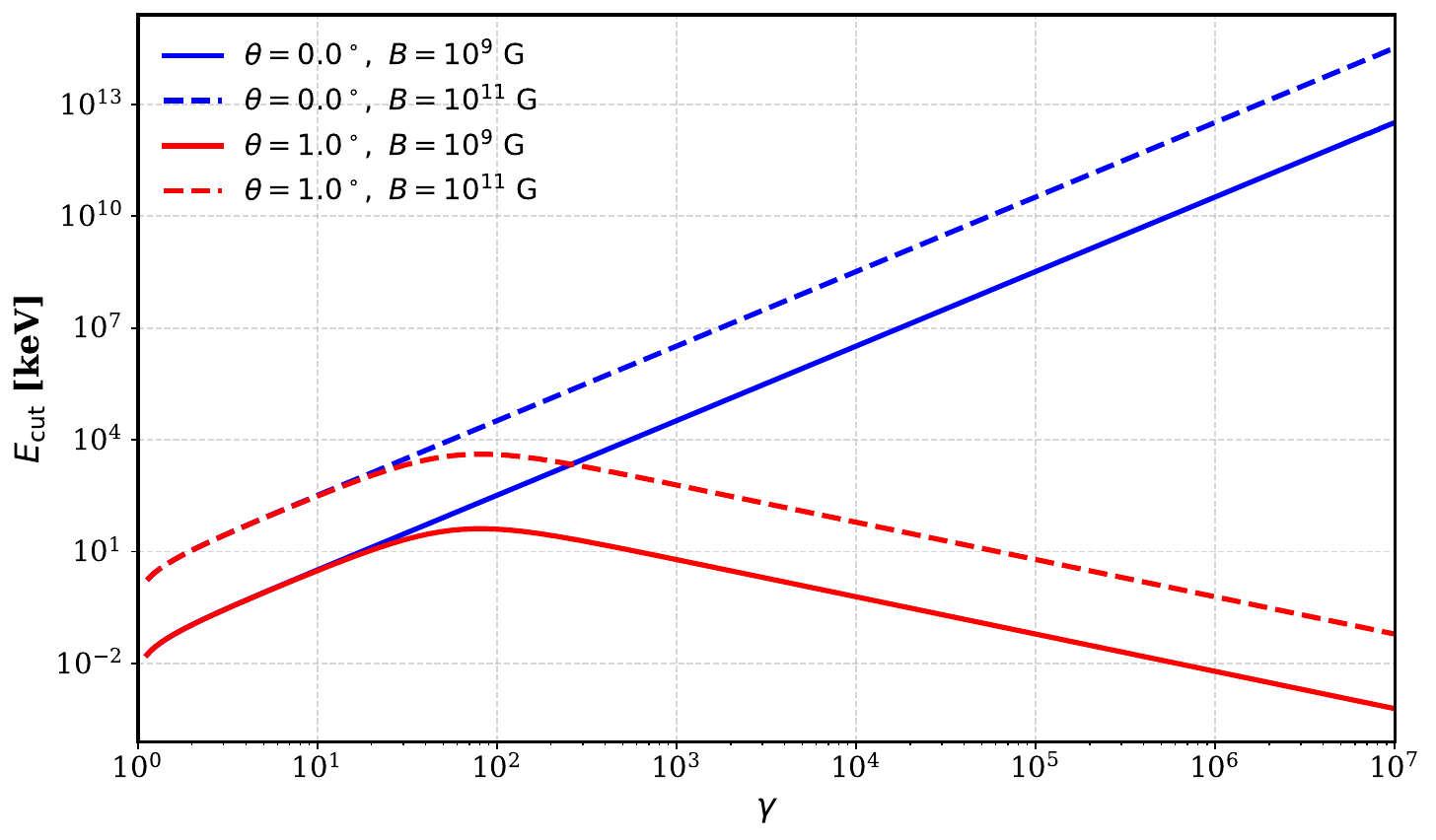}
    \caption{The relation between cutoff energy ($\hbar n_c \omega_B$) with $\gamma$ for different $\theta$ and $B$. Clearly, for a fixed cutoff energy, $\gamma$ and $B$ have anti-correlation for $\theta = 0^{\circ}$, whereas for higher latitudes (here, i.e., $1^{\circ}$), they are correlated after a certain $\gamma$ (See text for detailed explanation).}
    \label{fig: Ecut_vs_gamma}
\end{figure}

Here, it is important to discuss the differences in degeneracy in the parameter space from emission close to $\theta = 0^{\circ}$ and higher-latitude angles. Interestingly, from Figure \ref{fig: Ecut_vs_gamma}, we can see that for $\theta = 0^{\circ}$, to get the same cutoff energy ($\hbar n_c \omega_B$) if $B$ increases, $\gamma$ will decrease, since the cutoff harmonic ($n_c$) directly depends on $\gamma$ (i.e, for ultrarelativistic case, $n_c \sim \gamma^3$). However, when the emission is at a non-zero latitude, the relation between $B$ and $\gamma$ becomes more and more correlated after a certain velocity of particles (i.e., after a certain $\gamma$ in the figure). This can be easily understood from the cutoff frequency $\omega_c = n_c \omega_B$, where $n_c \simeq \frac{\beta'^{1/2}}{(1-\beta')^{3/2}}$, and $\beta' = \beta \cos\theta$. The fact is that after a certain $\gamma$ (or $\beta$), the harmonic is mostly dominated by $\theta$. So, cutoff energy $E_{cut} \propto B/\gamma$ (from $\omega_B$). For the $\theta=0^{\circ}$ case, $E_{cut} \propto \gamma^2 B$ as known for angle-averaged synchrotron emission from ultrarelativistic particles (see \citealt{Ghosh2023} for more details). 

This correlation between $\gamma$ and $B$ in the case of higher latitudes prevents us from measuring them separately by spectral fitting. Also, one can see that for a fixed $B$,  two different values of $\gamma$ give the same $E_{cut}$ for the $1^{\circ}$ angled emission (figure \ref{fig: Ecut_vs_gamma}), i.e, one is in the rising trend, where $\gamma$ and $B$ are anti-correlated, the other one is in the declining trend, where they are correlated. Similar trends apply to any high-latitude angle scenario. However, these two different $\gamma$ values will give two different spectral shapes, thus such degeneracy can be easily removed while comparing with the observed spectrum. The ULX spectra are best fitted with parameters from this correlated (declining trend) parameter space (see section \ref{Sec: Results}). Thus, with this prelude of the explained model, we move forward with spectral fitting and explaining the physical properties from real observations.

\section{Data analysis}\label{Sec: Data}
We proposed the first joint X-ray observation of NGC 470 HLX1 using \xmm\ and \nustar\ to comprehend its broadband X-ray coverage. The main goal of the paper is to explore the physics of synchrotron radiation in this source, along with the soft spectral component, and to comprehend the complete physical picture of the source properties. Hence, we primarily discuss the \xmm\ and \nustar\ data in detail (see Table \ref{tab:logtable}). The data have been extracted and analyzed using the standard method described below.

The \xmm\ data are processed using SASv22.1.0, and \texttt{epproc} and \texttt{emproc} tools are used to extract the EPIC products. The data are background flare-corrected, and the scientific products are extracted using \texttt{evselect} task following the standard procedure with filters PATTERN$<=$4 for pn and PATTERN$<=$12 for MOS. In spectral extraction, FLAG==0 is used for all CCDs. \texttt{rmfgen} and \texttt{arfgen} tasks are used to generate response and ancillary files, and the spectra are grouped with $20$ minimum counts per bin with an oversampling factor of $3$ using \texttt{specgroup}. The source region is a $15$ arcsec circle, and the background is a $30$ arcsec circle in a source-free region from the same chip. 

The \nustar\ data are processed with HEASOFT v6.35.1. \texttt{nupipeline} task is used to create the cleaned data with \texttt{saacalc=3}, \texttt{saamode=OPTIMIZED}, and \texttt{tentacle=yes} parameters to maintain a balanced approach while handling the background due to South Atlantic Anomaly and keeping the highest possible on-time exposure. \texttt{nuproducts} is used to extract spectra, response files, and light curves. The spectrum is grouped using a minimum of $20$ counts per bin. It is important to note that the distance between the source NGC 470 HLX1 and the galactic center/QSO is $\gtrsim 30$ arcsec, as can be separated in the \xmm\ data. In addition, we find that the flux of the galaxy center, especially beyond $\sim 3$ keV, is significantly lower (factor of $\sim 4$) than HLX1. Hence, for the scientific purpose of this paper, a selection of $25$ arcsec circular source and a background region of annulus between $35$ arcsec and $65$ arcsec in the \nustar\ data is found to be suitable. In that way, any contamination from that source in the HLX1 spectra is minimized.

\begin{deluxetable*}{cccccc}
\tablenum{1}
\tablecaption{Observation log of NGC 470 HLX1 \label{tab:logtable}}
\tablewidth{0pt}
\tablehead{
\colhead{Serial No.} & \colhead{Observation ID} & \colhead{Observation ID} & \colhead{Observation start date}  &
 \colhead{Spectral Exposure time (ks)}  \\
\nocolhead{} & \colhead{\xmm} &  \colhead{\nustar} &
\nocolhead{}  & \colhead{pn/MOS1/MOS2/FPMA/FPMB}
}
\startdata
1 & 0200780101 & - &  2004-01-24  & 4/4/4/-/-  \\
2 & 0601670101 & - &  2009-06-27 & 33/54/55/-/-  \\
3 & 0903440201 & 30801007002 &  2023-01-09/2023-01-08  & 46/61/67/119/117 \\
\enddata
\end{deluxetable*}

\section{Results}\label{Sec: Results}
\begin{figure}
\centering
    \includegraphics[width=1. \linewidth]{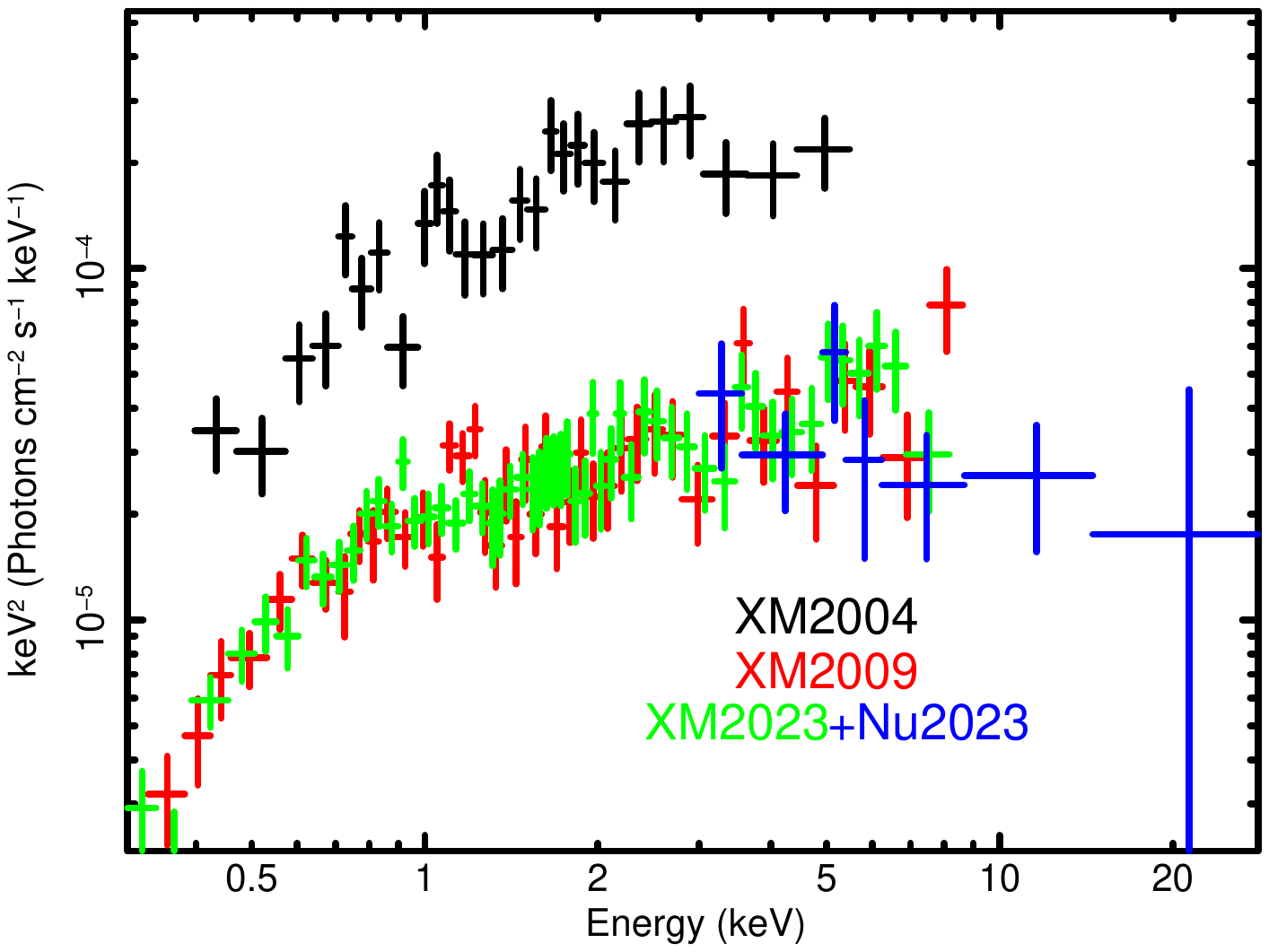}
\caption{The unfolded spectra are plotted using \texttt{powerlaw} model of 0 index and arbitrary normalization. For visual clarity, only \xmm\ pn and \nustar\ FPMA spectra are shown. The data have been rebinned for visual purposes.
\label{fig:spectra_eeufspec}}
\end{figure}

First, we analyze individual \xmm\ observations from 2004 and 2009, and the joint \xmm+\nustar\ observation from 2023. From Figure \ref{fig:spectra_eeufspec}, one can see that there is a significant flux variability from 2004 to 2009 observations. However, the 2009 spectral data overlap with those from the 2023 observation. When fitted with a simple absorbed \texttt{powerlaw} model to both 2009 and 2023 \xmm\ observations, the spectral index ($\sim 1.7$), and other parameters, i.e., the spectral shape, are found to be statistically overlapping (as also visually apparent from figure \ref{fig:spectra_eeufspec}). Hence, these two \xmm\ observations, including the \nustar\ data, were utilized for simultaneous fitting for further analysis to obtain better constraints on the parameters. The 2004 observation is analysed separately due to its significantly higher flux. For convenience, 2004, 2009, and 2023 \xmm\ data are denoted as XM2004, XM2009, and XM2023, respectively, and the \nustar\ data is denoted as Nu2023.

Throughout the paper, spectral analysis is performed using XSPECv12.15.0 \citep{XSPEC}, and we have quoted uncertainties on spectral parameters within a 90\% confidence interval unless otherwise specified. For modeling neutral absorption with \texttt{tbabs}, we employed parameters based on the elemental composition outlined by \cite{Wilms} and the absorption cross-sections provided by \cite{Verner}. Spectral data from \xmm\ were analyzed between 0.3 and 10.0 keV, while \nustar\ data were restricted to the 3.0 to 30.0 keV range, as background interference becomes substantial at higher energies.

We first use an absorbed \texttt{powerlaw} model to the simultaneous XM2009 and XM2023+Nu2023 spectral fit, which exhibits a powerlaw index of $\sim 1.74 \pm 0.07$, neutral absorption of $0.11 \pm 0.02 \times 10^{22}$ cm$^{-2}$ ($\chi^2/dof = 231/230)$. The neutral absorption value is in excess of the Galactic ISM absorption value ($\sim 0.03 \times 10^{22}$ cm$^{-2}$; \citealt{HI4PI}).

Since we have the \nustar\ coverage in the 2023 observation, we can detect the turnover in the spectrum. However, only an absorbed \texttt{cutoffpl} model does not provide a statistically improved fit. We find that in the softer spectral counterpart, an additional \texttt{blackbody (bb)} component improves the fit significantly $\chi^2/dof = 199/227$, with $N_H = 0.12^{+0.08}_{-0.06} \times 10^{22}$ cm$^{-2}$, $kT_{bb} = 0.18 \pm 0.04$ keV, $\Gamma \sim 0.9^{+0.4}_{-0.6}$, $E_{fold} \sim 5^{+6}_{-2}$ keV. We note that even though a simple absorbed \texttt{powerlaw} model provides an acceptable statistical fit, the investigation with \texttt{bb+cutoffpl} model and subsequent analysis with synchrotron model, is physically motivated, and also owing to the fact that a similar cutoff is observed in many ULX sources.

The XM2004 observation, is well fitted with a simple \texttt{powerlaw} component with $N_H = 0.23^{+0.08}_{-0.07} \times 10^{22}$ cm$^{-2}$, $\Gamma = 1.9 \pm 0.2$ ($\chi^2/dof = 46/41$). A \texttt{cutoffpl} model gives $\chi^2/dof = 42/41$, with $N_H$ fixed to the Galactic value (since free $N_H$ is unconstrained), $\Gamma \sim 0.5^{+0.3}_{-0.4}$, $E_{fold} \sim 1.9^{+0.9}_{-0.5}$ keV. Addition of a \texttt{blackbody} does not improve the statistics, nor are the parameters constrained.

As our main goal is to investigate the implications of the newly developed \texttt{synchrotron} radiation model in this source, we study the model to explain the high-energy spectral curvature in the spectra. As explained in section \ref{Sec: Synchrotron Model}, we have searched for the best fit $\gamma_{min}$ parameter while keeping the $B$ and $\theta$ parameters fixed. $\gamma_{max} = 1000 \gamma_{min}$ is used, although $\gamma_{max}$ beyond a few tens of $\gamma_{min}$ won't play much role (see \citealt{Ghosh2023} for details).

We analyse the data with a range of $\theta$ and $B$ parameters, to capture several plausible scenarios, and their implications are discussed in the next section. It is important to note that knowing the exact $\theta$ is not possible unless we can certainly determine the $B$ or the $\gamma$ parameter from any other physical estimates. Thus, we choose some fixed values of $\theta$, and $B$, to get a general idea of the emission mechanism.

We find that, in addition to the synchrotron component, a blackbody component is required to get a good statistical fit to the broadband data (2009 and 2023 epochs). Thus, the complete continuum that we fit is \texttt{bb+synchrotron}. The parameters are detailed in Table \ref{tab:sync_table}. We find that for 1$^{\circ}$ angle emission, if the magnetic field is around $10^9$ G, then the $\gamma_{min}$ is around $\sim 1000$. However, for $10^{12}$ G, we will have a much larger $\gamma_{min} \sim 10^6$ to explain the cutoff. For 30$^{\circ}$ latitude emission, for $10^{12}$ G field, $\gamma_{min}$ is around $\sim 30$, whereas a field as low as $10^9$ G would not produce photons with $10$ keV cutoff at such a high latitude. It is important to note that for higher magnetic fields, a wide range of latitudes (from low to high latitudes) can play a role in explaining the observed spectra. The lower the magnetic field strength, the more constricted the latitude, and it is restricted to a lower emission angle to explain the observed spectral cutoff.  

\begin{figure*}[t]
\centering
    \includegraphics[width=\columnwidth]{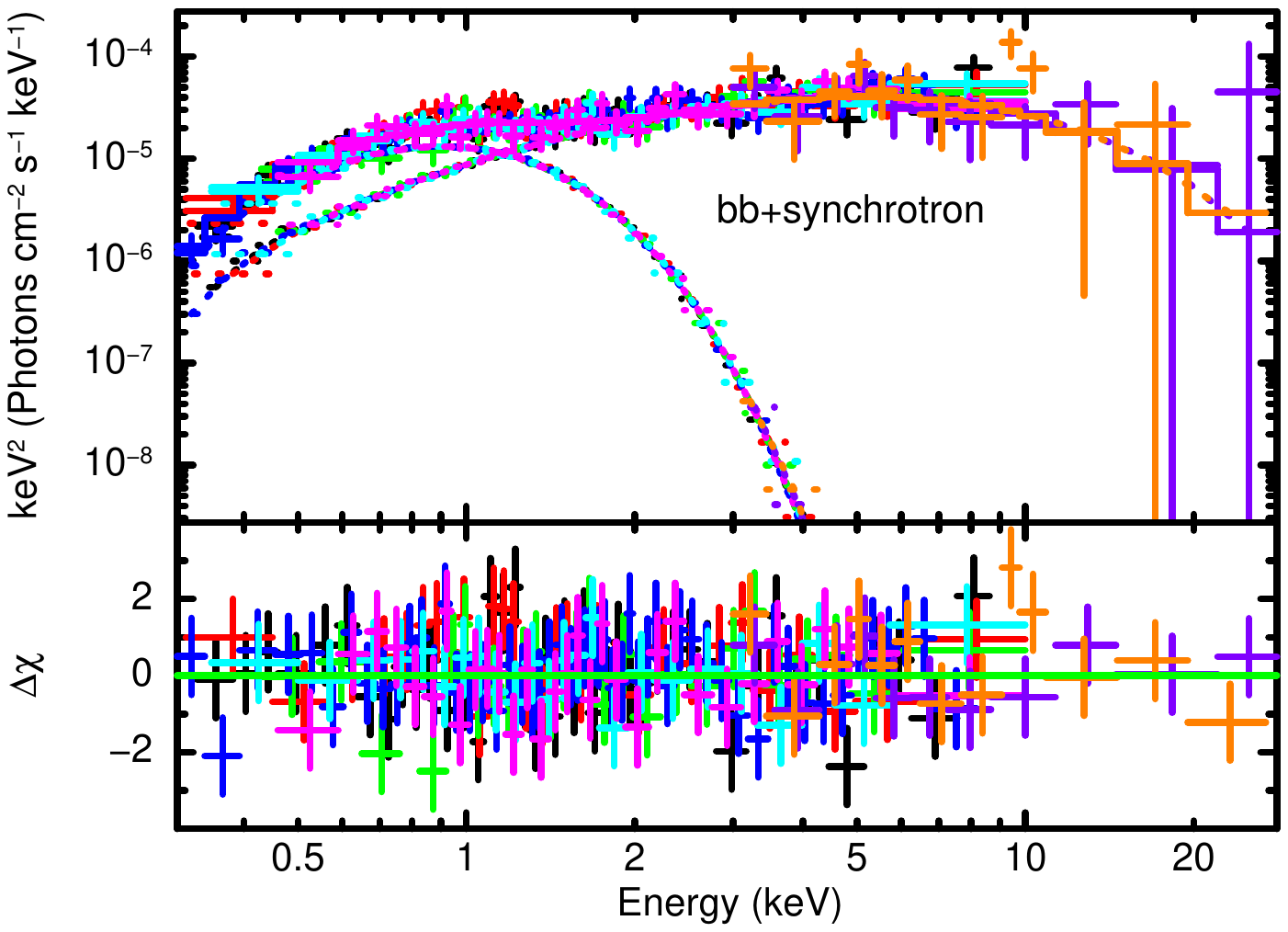}
    \includegraphics[width=\columnwidth]{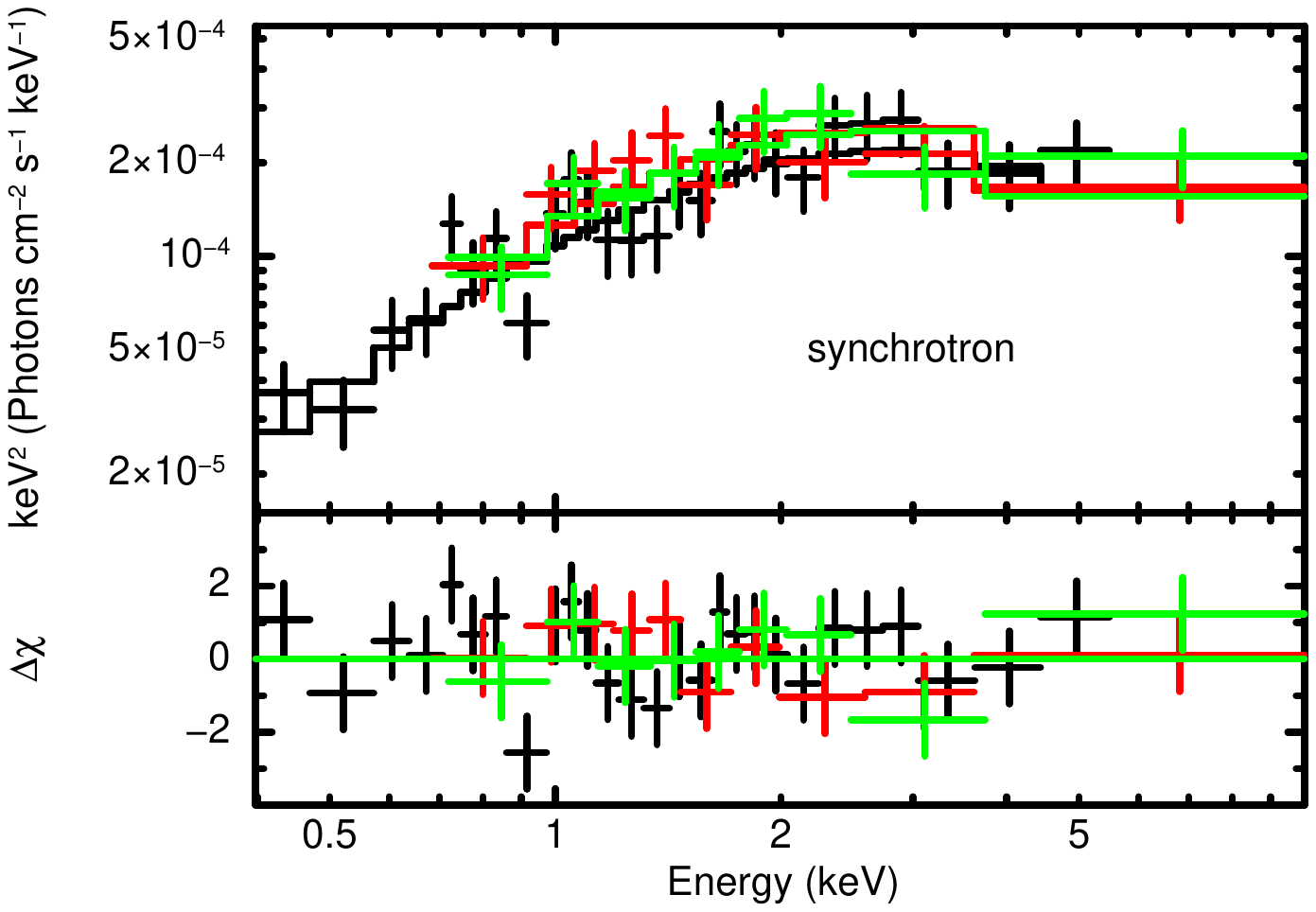}
\caption{Left: The spectra, model components, and residual for \texttt{bb+synchrotron} fit for broadband data (XM2009 and XM2023+Nu2023). Right: Same for the XM2004 data with only \texttt{synchrotron} model.
\label{fig:residual_plot_bb}}
\end{figure*}

We see that the absorption $N_H \sim 0.08 \times 10^{22}$ cm$^{-2}$, and blackbody parameters (with $T \sim 0.2 $ keV) remain unchanged for different parameter combinations of \texttt{synchrotron} model. The $0.3\hbox{--}30.0$ keV unabsorbed flux in this epoch is estimated to be $1.7 \pm 0.1 \times 10^{-13}$ \fluxcgs, corresponding to luminosity of $L_x \sim 2.35 \pm 0.15 \times 10^{40}$ \lumcgs. Corresponding spectra and residuals are shown in Figure \ref{fig:residual_plot_bb} -Left for the broadband fit.

For the high flux state of the source, i.e., XM2004 observation, we find that a single component of synchrotron model without the need of blackbody component can fit the data ($\chi^2/dof = 43/42$), however, predicting a lower spectral cutoff energy, i.e, $\gamma_{min} = 63 \pm 6$ for a $10^{12}$ G field, and $\theta = 30^{\circ}$, and a higher normalization of $N_{sync} = 89^{+13}_{-12} \times 10^{-20}$ cm$^{-2}$. Here, we also had to fix the $N_H$ to the Galactic ISM absorption value since it was unconstrained as a free parameter. The trend of synchrotron parameters remains the same for all other combinations, i.e., $\gamma_{min}$ is approximately twice given the same $\theta$ and $B$, compared to the other epochs of the source. The normalization of synchrotron is $\sim 20$ times higher compared to the low flux state. Nevertheless, the acceptable statistical fit with the synchrotron model in the high flux state of the source quantitatively validates this non-thermal component in different flux states.  The $0.3\hbox{--}30.0$ keV unabsorbed flux in this high flux epoch is estimated to be $7.0^{+0.9}_{-0.6} \times 10^{-13}$ \fluxcgs, corresponding to luminosity of $L_x \sim 9.7 \pm 1.0 \times 10^{40}$ \lumcgs. Corresponding spectra and residuals are shown in Figure \ref{fig:residual_plot_bb} -Right for the XM2004 fit.

Apart from spectral analysis, we explore the time series from \xmm\ and \nustar\ data of NGC 470 HLX1 and find that there is no significant short-term variability in the light curves. We search for pulsation in the source using the acceleration search technique by \texttt{HENaccelsearch} and \texttt{HENzsearch} task of the \texttt{HENDRICS} \citep{HENDRICS, Stingray} package in $0.1\hbox{--}10.0$ Hz for \xmm\ pn
large window mode data (XM2023 observation), and \nustar\ data, and within $0.1\hbox{--}6.8$ Hz for \xmm\ pn full frame mode observations (XM2004 and XM2009). No significant detection of any pulsed signal can be comprehended from our search.

\begin{deluxetable*}{cccccccc}
\tablenum{2}
\tablecaption{Parameters of \texttt{bb+synchrotron} model for different $\theta$ and $B$ combinations for broadband spectral fitting of 2009 and 2023 observations. \label{tab:sync_table}}
\tablewidth{0pt}
\tablehead{
\colhead{$\theta$} &
\colhead{$N_H \,(10^{22}\,\mathrm{cm}^{-2})$} &
\colhead{$T_{bb}$ (keV)} &
\colhead{$Norm_{bb}\,(10^{-7})$} &
\colhead{$\gamma_{min}$} &
\colhead{$B$ (G)} &
\colhead{$N_{sync}\,(10^{-20}\,\mathrm{cm}^{-2})$} &
\colhead{$\chi^2/\mathrm{dof}$}
}
\startdata
$0.5^{\circ}$ & $0.08^{+0.05}_{-0.04}$ & $0.21 \pm 0.03$ & $4.16^{+1.24}_{-0.79}$ & $7.5^{+0.9}_{-0.8} \times 10^6$ & $10^{12}$ & $357^{+76}_{-66}$ & $202/228$ \\
$0.5^{\circ}$ & $0.08^{+0.05}_{-0.04}$ & $0.21 \pm 0.03$ & $4.15^{+1.23}_{-0.78}$ & $7.4^{+1.0}_{-0.8} \times 10^3$ & $10^{9}$ & $353^{+81}_{-61}$ & $202/228$ \\
\hline
$1^{\circ}$ & $0.08^{+0.05}_{-0.04}$ & $0.21 \pm 0.03$ & $4.15^{+1.19}_{-0.78}$ & $9.3^{+1.2}_{-1.0} \times 10^5$ & $10^{12}$ & $177^{+40}_{-31}$ & $202/228$ \\
$1^{\circ}$ & $0.08^{+0.05}_{-0.04}$ & $0.21 \pm 0.03$ & $4.15^{+1.21}_{-0.78}$ & $9.3^{+1.2}_{-1.1} \times 10^2$ & $10^{9}$ & $178^{+38}_{-32}$ & $202/228$ \\
\hline
$30^{\circ}$ & $0.08^{+0.05}_{-0.04}$ & $0.21 \pm 0.03$ & $4.19^{+1.19}_{-0.76}$ & $34^{+5}_{-4}$ & $10^{12}$ & $5 \pm 1$ & $202/228$ \\
$30^{\circ}$ & - & - & - & - & $10^9$ & - & - \\
\enddata
\end{deluxetable*}

\section{Discussions} \label{sec:discussions}

We study one of the extremely luminous sources that has exhibited luminosity in the HLX range ($>10^{41}$ \lumcgs; \citealt{Walton2011, Sutton2012}), and investigate its spectral properties from the perspective of the newly developed non-thermal incoherent latitude-dependent synchrotron emission model\citep{Ghosh2023}. 

The apparent spectral cutoff confirmed by the new broadband data, especially thanks to \nustar, can be described by this synchrotron model. Also, we find that a soft blackbody component is needed to describe the broadband data. The XM2004 observation did not require an additional blackbody component; however, that could be due to its low signal-to-noise ratio (S/N) and count statistics because of limited clean on-time exposure. Another argument could be simply because the spectral folding was detected with a comparatively lower energy, and one single \texttt{cutoffpl} component is sufficient to model the spectra. Nevertheless, the feasibility of the synchrotron model is tested in both low-flux and high-flux states of the source. During the low-flux state, the source exhibits a few factors of lower flux compared to the high-flux state, but it can still be considered as an extremely luminous ULX regime. Interestingly, a similar flux state change has been observed in one of the brightest PULXs, NGC 5907 ULX1 \citep{Furst2023}. Therefore, growing evidences suggest that HLXs and extremely luminous ULXs could also be powered by neutron stars. In that scenario, it is promising that a synchrotron emission model, possibly coming from a neutron star magnetic field strength, explains the spectral feature of NGC 470 HLX1, making it another possible NS candidate. Nevertheless, without concrete evidence of pulsation, it would be difficult to conclude its nature.

The parameters of the synchrotron model give an insight that for a fixed magnetic field, the more relativistic plasma would have lower latitude emission to explain the observed spectra, and vice versa (see also \citealt{Ghosh2023}). Also, let us consider the example parameter space chosen in our fitting; the normalization, which is related to the number density of particles as $\rho V/D^2$, where $V$ is the emission region volume, would estimate relativistic particle number density ($\rho$) responsible for the emission around $10^{11}\hbox{--}10^{16} $ cm$^{-3}$, if we assume typical emission region volume within the magnetosphere. However, it is important to note that this number density will be uncertain unless the exact location of the particle acceleration of these sources can be precisely determined.  

Theoretically, we expect that acceleration of electrons will not be significant inside the accretion column for a high surface $B$ field source, since the synchrotron cooling will not allow the electrons to achieve very high $\gamma$ inside the column; however, that might be the relevant case of a high latitude and high magnetic field scenario in our model.

Nevertheless, different lepton acceleration method suggests that electrons can indeed get significantly accelerated outside the column till the magnetosphere region. Therefore, the magnetic field from which synchrotron radiation can originate would possibly have a broad range, which needs to be taken into account, depending on the surface magnetic field of the source. Also, we can extend our understanding from the analysis that for lower magnetic field strengths (i.e., $10^9$~G), high latitude emission will not provide $\sim 10$ keV spectral cutoff. For such a low magnetic field, only emission with latitude close to the orbital plane of the electrons would give the observed spectral cutoff. However, a higher magnetic field at any location of the magnetosphere can provide such a cutoff with a higher latitude, as mentioned in our analysis. 

Several electron acceleration mechanisms have been explored relevant to accreting neutron star systems, such as collisional shock acceleration due to wind from the companion onto the Alfven radius of the neutron star \citep{Bednarek2009}. There, the acceleration happens close to the boundary of the magnetosphere and the stellar wind due to a shock, parameterized by the acceleration parameter and Larmor radius of the electrons (also related to the Hillas criterion \citealt{Hillas1984}). Also, different particle acceleration methods can be responsible for a swarm of relativistic electrons in the plasma, and it can occur at different locations of the magnetosphere. 

\begin{figure}
    \centering
    \includegraphics[width=\linewidth]{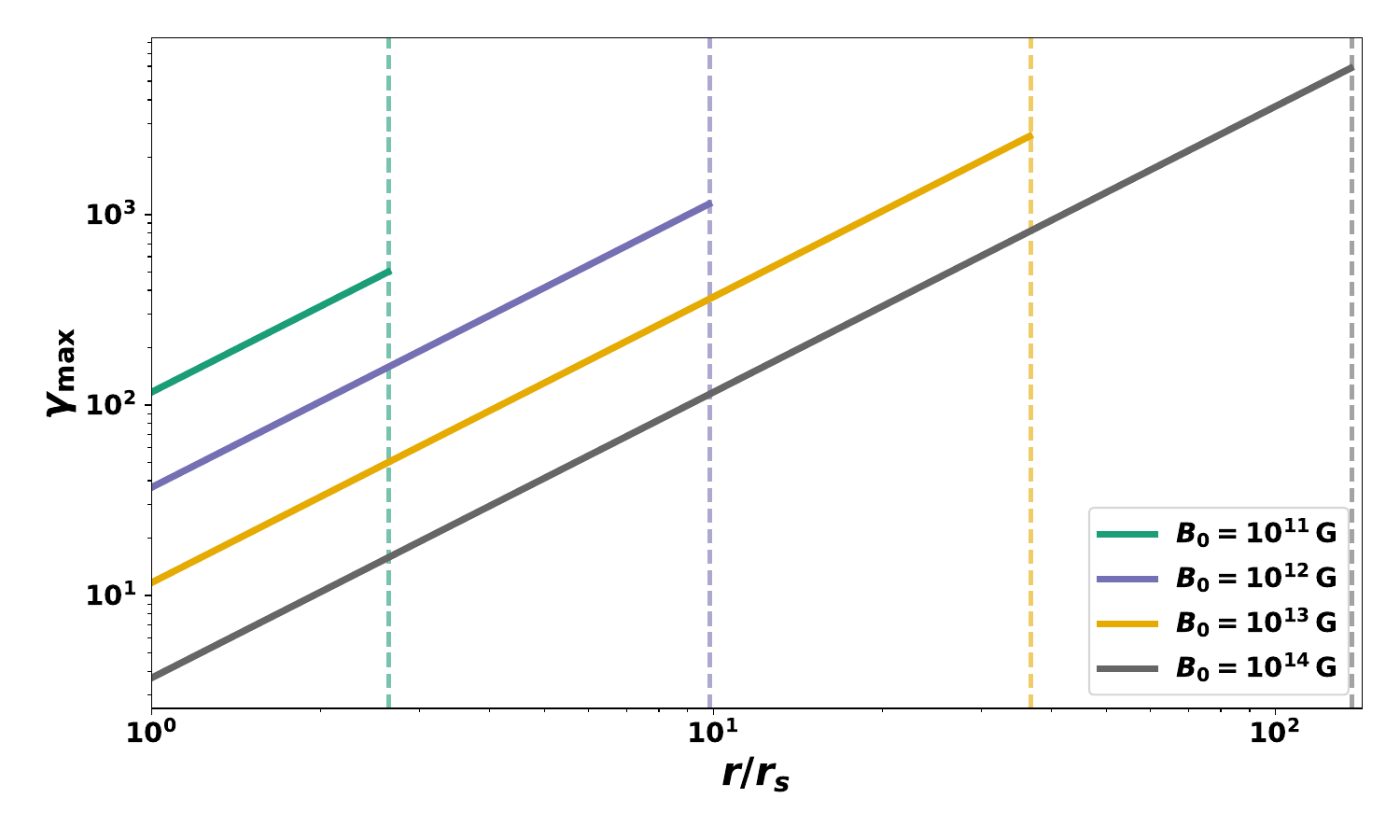}
    \caption{Electron maximum Lorentz factor varying with distance from the NS surface, by balancing energy gains from the acceleration process with synchrotron cooling losses for different surface magnetic field strengths $B_0$ (see text for details). The vertical dotted lines show the corresponding $R_M$ for each $B_0$ case.}
    \label{fig: Hillas}
\end{figure}

If we consider the simple scenario of particle acceleration parametrized by the Larmor radius and acceleration parameter and consider synchrotron cooling as the dominant cooling mechanism within the magnetosphere (see also \citealt{Bednarek2009}), we can see how the electron's maximum Lorentz factor would vary over distance from the surface till the magnetospheric radius $R_M = 7\times 10^7 \Lambda m^{1/7} R_6^{10/7} B_{12}^{4/7} L_{39}^{-2/7}$ cm (see details in \citealt{Mushtukov2017, GhoshNGC6946}). To calculate the $R_M$, we take luminosity around $\approx 10^{41}$ \lumcgs for the source. The energy gain rate due to acceleration is typically $\dot E_{gain} = \eta e B c$, where $\eta \sim 0.1$ is acceleration parameter \citep{Bednarek2009}. The energy loss rate due to synchrotron is $\dot E_{loss} = (4/3) c \sigma_T U_B \gamma^2$. Thus, after correcting for the cooling effects, the $\gamma_{max} = \sqrt \frac{3 \eta e B}{4 \sigma_T U_B}$. Here $U_B = \frac{B^2}{8\pi}$. See figure \ref{fig: Hillas} for the variation of maximum Lorentz factor with distance from the surface.

There are other processes as well, like large-amplitude Deutsch waves \citep{Gunn1969, Eichler1984}, acceleration via electromotive forces, and magnetic reconnection due to strong conduction currents in the disk-magnetosphere-star circuit \citep{Miller1994, Hamilton1994}, and magnetospheric gap induced acceleration \citep{Cheng1989, Cheng1991, Zhang2014}. In all of these acceleration mechanisms, it is argued that electrons can not have an energy larger than $10^{13}$ eV, i.e., giving an upper limit of the Lorentz factor as $\gamma_{max} \sim 10^7$. Based on those acceleration mechanisms, the feasibility of a non-thermal synchrotron from electrons is evident in accreting neutron star systems.

The synchrotron radiation could emerge from either the accreting column close to the neutron star surface or the magnetosphere. One important criterion for observing this radiation is that it doesn't share substantial energy with the surrounding medium through inverse Compton scattering. The number density of non-relativistic particles in the accreting column is expected to be $n_{\rm nr} \gtrsim 10^{21} \, \rm cm^{-3}$, its height to be 10~kms, and its width to be around 1~km. In our proposed scenario,
the photon emerges at an angle to the column. The fractional  energy exchange between the X-ray photon and the medium  owing to inverse Compton scattering is characterized by the y-parameter: 
$y = n_{\rm nr} \sigma_T l \delta \epsilon/\epsilon$ where $\delta \epsilon/\epsilon = (4KT-\epsilon)/(m_e c^2)$ and $l$ is the path traversed by the photon in the medium. If $y \ll 1$, we do not expect much energy exchange between the X-ray photon and the surrounding medium. This condition is readily met in the magnetosphere with much lower $n_{\rm nr}$. It could also be met close to the accreting column, e.g., if $n_{\rm nr} \simeq 10^{21}  \, \rm cm^{-3}$, $l \simeq 10^5 \, \rm cm$. 

A simple inspection to check whether the emission is synchrotron or Compton dominated is to compare the magnetic energy density $U_B = \frac{B^2}{8\pi}$ and radiation energy density $U_{rad} = L/A c$, where $A$ is the area of the local region of interest and $L$ is the luminosity. It is straightforward to see that $U_B$ dominates within the magnetosphere from close to the surface upto the magnetospheric radius ($R_M$) in these systems, hence synchrotron emission dominates. We also find that in the relevant X-ray photon energy ranges, the medium is optically thin to the synchrotron self-absorption process \citep{Rybicki1979, Warren2018}.

Nevertheless, having discussed these approximate estimates of particle acceleration and efficiencies, it is crucial to investigate in the future with a robust simulated environment to understand how these processes evolve within a super-Eddington accretion scenario.

Regarding the blackbody component, the origin of such soft emission still requires a concrete physical explanation in the context of ULXs, which is beyond the scope of this paper. Here, for example, the luminosity of the blackbody component is much higher than what is expected from a polar cap emission with the observed temperature. Therefore, as predicted for many other ULX sources, the soft blackbody component could be related to the disk close to the corotation radius or outflow ejected from the disk \citep{Pintore2017}.

In this paper, we shed light on the feasibility of synchrotron emission and the overall broadband spectral model for NGC 470 HLX1, which shows a possibility that this source could be a neutron star. Similarly, the explanation would hold for other ULXs with such a typical spectral cutoff (see e.g., \citealt{Ghosh2023}). 

Although, as discussed before, we cannot distinguish the different latitude emissions from the current spectral fitting. However, we note that very high-latitude emission for a fixed magnetic field would require a semi-relativistic plasma and would produce detectable discrete cyclotron lines and elliptically polarized photons, which could be a critical point in distinguishing different latitudes in future observations (see also \citealt{Ghosh2023}). In fact, we can also distinguish from future polarization measurement whether the radiation is dominated by synchrotron or other mechanisms, e.g., Compton scattering. 

Additionally, with an understanding of the acceleration processes of electrons, as depicted in a simplified picture in Figure \ref{fig: Hillas}, we can perceive the typical range of Lorentz factors, which in turn constrain the most plausible values of $\theta$ and $B$ from our analysis. For example, if very high Lorentz factors are not plausible in the system, we can rule out some of the parameter space from our analysis (see table \ref{tab:sync_table}), i.e., low-angled emission in the presence of high magnetic fields. However, to exactly determine all the parameters, we need to know some of them from other theoretical arguments or analysis.

Moreover, we can further understand that for such an extremely luminous system, in the presence of a very high accretion rate, a strong surface magnetic field would be required to sustain the magnetosphere outside the neutron star surface. In that case, from the figure \ref{fig: Hillas}, we can predict the location of synchrotron emission. For a $B_0 \sim 10^{11}$~G field, even close to the surface, particles can achieve a reasonable lorentz factor for synchrotron emission, however, stronger the surface field strength (e.g., $B_0 \sim 10^{14}$ ~G), the location of the emission should be farther away from the neutron star surface; since, close to the surface of the neutron star, particles will not be highly relativistic to efficiently produce synchrotron emission in such a scenario.

Another aspect of the emission mechanism in these systems, which needs caution to understand, is the total available power. This is indeed a crucial problem in ULXs and HLXs, especially in accreting neutron star systems, which raises the question of how such a significant power budget is achieved. In the future, a comprehensive treatment is warranted to investigate this issue further in the context of synchrotron radiation and the maximum power of accelerated electrons.

This model provides a new window to look at the high-energy emissions from accreting neutron star X-ray binary systems. An important follow-up, which ought to be carried out, is to investigate the feasibility of this model on high S/N Galactic neutron star XRB data, which shows a similar cutoff in the hard X-ray energy range (e.g., Cen X-4; \citealt{Chakrabarty2014}). That will provide us with leverage to constrain the parameters precisely, since several other observational evidences will help us know some model parameters a priori (e.g., magnetic field strength), thus constraining other parameters more precisely.

In the future, it is prudent to analyse a sample of HLX sources and investigate whether a neutron star can power such high luminosity in this class of sources. In that way, the existing consensus of ``many HLXs are most likely IMBHs" can be challenged, and perhaps ``many HLXs are stellar mass super-Eddington neutron star accretors like ULXs" can be argued to be stronger.

\begin{acknowledgments}
We would like to thank the referee for the valuable suggestions that contributed to further improving the manuscript. In this work, we have used data (available at the High Energy Astrophysics Science Archive Research Center (HEASARC)) obtained with \xmm, an ESA science mission with instruments and contributions directly funded by ESA member states, and NASA. We have also utilized data (HEASARC) obtained with \nustar, a project led by Caltech, funded by NASA, and managed by the NASA Jet Propulsion Laboratory (JPL), and has made use of the NuSTAR Data Analysis Software (NuSTARDAS) jointly developed by the ASI Space Science Data Centre (SSDC, Italy) and the California Institute of Technology (Caltech, USA).
\end{acknowledgments}





%
\facilities{\xmm; \citet{XMM2001}, \nustar ; \citealt{NuSTAR} }

\software{HEASOFT (\url{https://heasarc.gsfc.nasa.gov/docs/software/heasoft/}; \citet{Heasoft2014}), \xmm\ SAS (\url{https://www.cosmos.esa.int/web/xmm-newton/sas}; \citet{Gabriel2004}), HENDRICS (\url{https://hendrics.stingray.science/en/latest/}; \citet{HENDRICS}), STINGRAY (\url{https://docs.stingray.science}; \citet{Stingray})
          }




\bibliography{NGC470HLX1}{}
\bibliographystyle{aasjournalv7}



\end{document}